\documentclass[prl,11pt, reprint, superscriptaddress]{revtex4-1}

\usepackage{graphicx}
\graphicspath{ {figures/} }
\usepackage{float}
\usepackage{hyperref}

\begin{document}

\title{Order from chaos: Observation of large-scale flow from turbulence in a two-dimensional superfluid}
\author{Shaun P.\ Johnstone}
\affiliation{School of Physics and Astronomy, Monash University, Victoria 3800, Australia}
\author{Andrew J.\ Groszek}
\affiliation{School of Physics and Astronomy, Monash University, Victoria 3800, Australia}
\author{Philip T.\ Starkey}
\affiliation{School of Physics and Astronomy, Monash University, Victoria 3800, Australia}
\author{Christopher J.\ Billington}
\affiliation{School of Physics and Astronomy, Monash University, Victoria 3800, Australia}
\affiliation{Joint Quantum Institute, National Institute of Standards and Technology, and University of Maryland, Gaithersburg, Maryland, 20899, USA}
\author{Tapio P.\ Simula}
\affiliation{School of Physics and Astronomy, Monash University, Victoria 3800, Australia}
\author{Kristian Helmerson}
\affiliation{School of Physics and Astronomy, Monash University, Victoria 3800, Australia}
\affiliation{ARC Centre of Excellence in Future Low-Energy Electronics Technologies, Monash University, Victoria 3800, Australia}

\date{\today}

\begin{abstract}
Interacting systems driven far from equilibrium tend to evolve to steady states exhibiting large-scale structure and order.
In two-dimensional turbulent flow the seemingly random swirling motion of a fluid can evolve towards persistent large-scale vortices.
Lars Onsager proposed a model based on statistical mechanics of quantized vortices to explain such behavior.
Here we report the first experimental confirmation of Onsager's model of turbulence.
We drag a grid barrier through an oblate superfluid Bose--Einstein condensate to generate non-equilibrium distributions of vortices.
We observe an inverse energy cascade driven by the evaporative heating of vortices, leading to steady-state configurations characterized by negative temperatures.
Our results open a pathway for quantitative studies of emergent structures in interacting quantum systems driven far from equilibrium.
\end{abstract}

\maketitle

One of the triumphs of 20$^{th}$ century physics was the development of statistical mechanics, allowing the description of emergent, collective behavior of many-particle systems near equilibrium.
However, much of the world around us is the result of systems far from equilibrium, for which a complete theoretical statistical description remains elusive~\cite{noauthor_condensed-matter_2007}.
Far from equilibrium behavior is apparent in hydrodynamic turbulence where the kinetic energy of the fluid is transported without loss across many length scales.
Yet, despite the complexity of the situation, there are some statistical theories that describe the steady state behavior of turbulent systems, such as Kolmogorov's power-law scaling of energy flow~\cite{kolmogorov_local_1941}.
In three-dimensional turbulence a process known as a Richardson cascade~\cite{richardson_weather_1922} results in energy transport to ever smaller length scales, causing vortices to break up over time and the system to appear chaotic.
Remarkably, restricting the fluid dynamics to two dimensions (2D) results in an inverse cascade process: energy flows towards the largest length scales available, resulting in large-scale, persistent vortex flows~\cite{kraichnan_inertial_1967}.
This behavior has been observed in systems ranging in scale from soap films~\cite{kellay_two-dimensional_2002} to Jupiter's atmosphere~\cite{young_forward_2017,adriani_clusters_2018}.

Onsager proposed an explanation for the appearance of large-scale vortex flow in two-dimensional turbulence in terms of equilibrium statistical mechanics of a model of quantized point vortices~\cite{onsager_statistical_1949}, noting its applicability to superfluids. 
He showed that the steady-state behavior of the model is described by a Boltzmann distribution with negative temperatures.
These negative temperature states correspond to high energy, highly ordered states, dominated by the large scale vortices, such as those produced by an inverse energy cascade in 2D.
Negative absolute temperature states were subsequently used to explain observations in nuclear spin systems~\cite{purcell_nuclear_1951} and, more recently, of motional degrees of freedom of cold atoms in optical lattices~\cite{braun_negative_2013}.
In these experiments, changing spin states, and the sign of inter-atom interactions, respectively, resulted in a forced population inversion with the resulting state decaying to lower energy while increasing its entropy.
However, until now, negative absolute temperature states have not been realized in the context of bounded systems of quantized vortices, as originally predicted by Onsager.
This vortex system is markedly different from the nuclear spin and optical lattice experiments in that, as opposed to being essentially a change of state for single particles, here energy is injected into the system in a continuous manner and the interactions of the constituent particles (vortices) result in a negative temperature configuration.

In a superfluid, three-dimensional quantum turbulence (3DQT), which manifests as tangles of quantized vortex lines, has been shown to exhibit a similar direct energy cascade to its classical counterpart.
The statistical dynamics of 3DQT have been studied over the past three decades, both numerically, and experimentally in superfluid helium~\cite{tsubota_numerical_2017}.
More recently 3DQT has been observed in atomic BECs~\cite{henn_emergence_2009,navon_emergence_2016}, where direct imaging is possible due to the comparatively large vortex cores (micron size vs.\ the {\AA}ngstr{\"o}m size in superfluid helium).
In two-dimensional quantum turbulence (2DQT) an inverse energy cascade is predicted to result from the preferential transport of the energy injected into the superfluid through the creation of vortex-antivortex pairs~\cite{reeves_inverse_2013,simula_emergence_2014}.
To date, however, the major challenge hindering the understanding of 2DQT has been to devise a method to experimentally measure the velocity field of a superfluid.

\begin{figure*}[ht]
    \centering
    \includegraphics[]{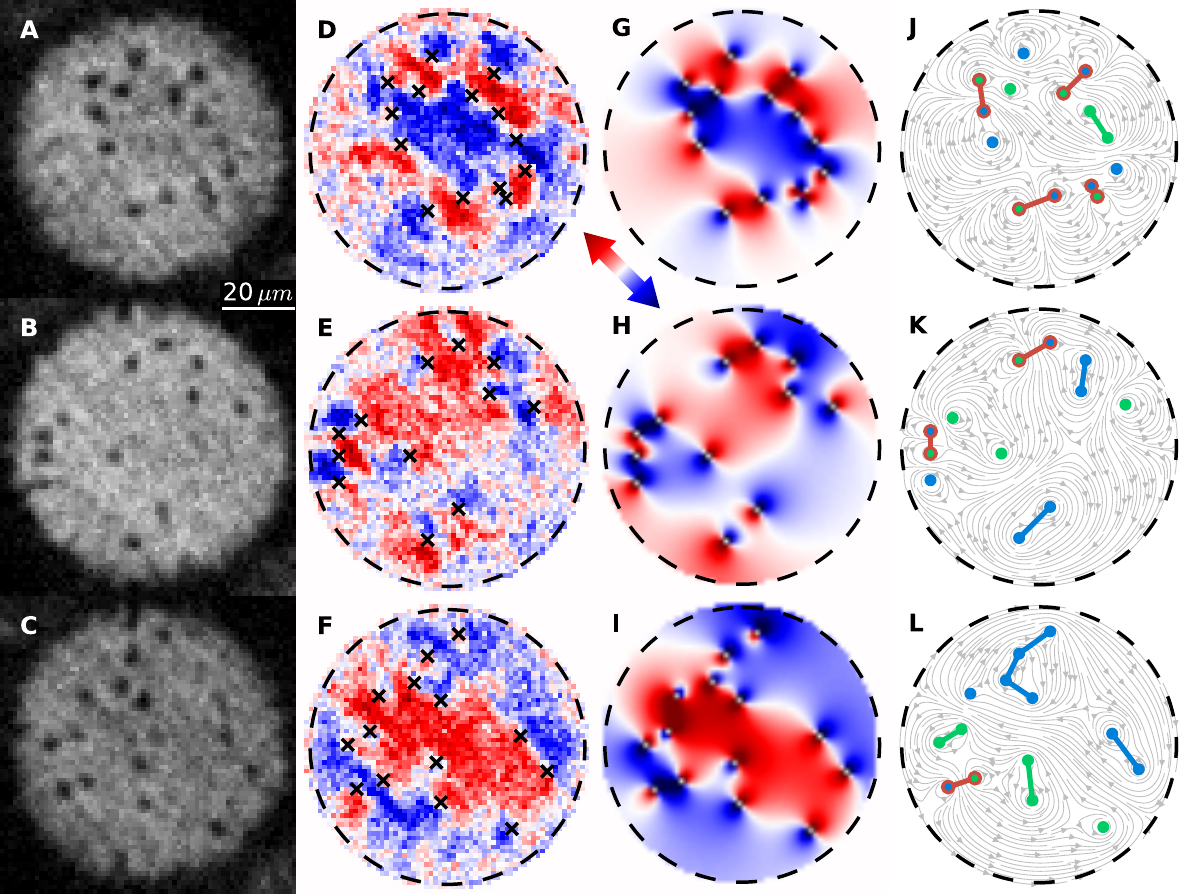}
    \caption{\textbf{Vortex configurations in the dipole, random and clustered regimes.}
    (\textbf{A} to \textbf{C}) The locations of vortices are visible as dark spots in the optical density images of the BEC.
    Here we show example distributions which are dipole dominated, random and cluster dominated, respectively.
    (\textbf{D} to \textbf{F}) The corresponding Bragg spectroscopy signals.
    (\textbf{G} to \textbf{I}) The computed velocity field projected onto the line defined by the directions of the Bragg spectroscopy laser beams.
    Colors in (D to I) indicate projections of the superfluid flow in the direction indicated by the arrow.
    (\textbf{J} to \textbf{L}) The classification of the vortices based on their signs and positions:
    vortices (antivortices) are indicated by blue (green) points;
    clusters by lines of the same color;
    dipoles are linked by red lines.
    Streamlines of the computed flow are shown in gray.
    }
    \label{fig:example_images}
\end{figure*}

Atomic Bose--Einstein condensates (BECs) provide an ideal system in which 2DQT can be realized, as they can be readily trapped in highly oblate geometries, where the dynamics of the vortices are restricted to a plane~\cite{neely_observation_2010,neely_characteristics_2013,kwon_relaxation_2014,seo_observation_2017}.
The emergence of persistent currents has been observed in an annular BEC~\cite{neely_characteristics_2013}, and the relaxation of turbulence has been investigated via vortex number statistics~\cite{kwon_relaxation_2014}.
These early experiments relied heavily on comparisons to numerical simulations, as information about the vortex circulation could not be obtained.
This provided motivation for developing new techniques to probe 2DQT, such as multi-shot vortex tracking~\cite{freilich_real-time_2010,wilson_situ_2015}, and single-shot vortex sign detection~\cite{powis_vortex_2014,seo_observation_2017}.
In the latter case, a velocity-selective Bragg spectroscopy~\cite{stenger_bragg_1999} technique was utilized, allowing the sign and position of each vortex (and hence the full velocity field) in a turbulent BEC to be determined for the first time~\cite{seo_observation_2017}.
However, the BEC was harmonically trapped and therefore inhomogeneous, and the vortices preferentially formed vortex-antivortex dipole pairs.
Indeed, numerical studies have indicated that the uniformity of the BEC plays a major role in the turbulent dynamics~\cite{groszek_onsager_2016}.

\begin{figure}[ht]
    \includegraphics[]{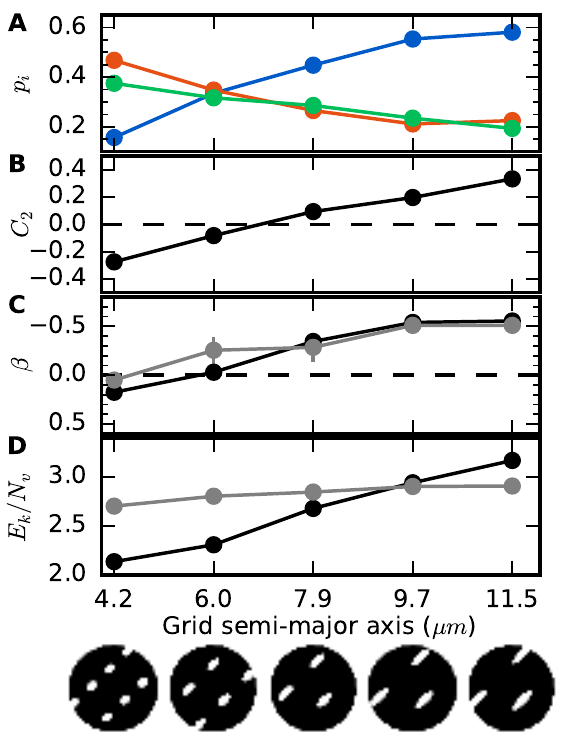}
    \centering
    \caption{\textbf{Generating grid turbulence at positive and negative absolute temperatures.}
        Vortex configuration data, time-averaged across early hold times after the grids have passed through the BEC, for each grid used (bottom).
		The grid configurations used are shown one third of the way through the 0.5\,s sweep.
        White areas have high laser intensity, repelling the atoms.
        (\textbf{A}) Classified vortex populations of clustered (blue), dipole (red) and free (green) vortices.
        (\textbf{B}) Correlation function $C_2$~\cite{methods}.
		(\textbf{C}) Inverse temperature of the vortices, $\beta$. Positive (negative) values are scaled by the critical temperature $|\beta_\mathrm{BKT}|$ ($|\beta_\mathrm{EBC}|$)~\cite{methods}.       
        (\textbf{D}) Incompressible kinetic energy (in units of $\rho \kappa^2/4\pi$, where $\rho$ is the superfluid density and $\kappa=h/m$ is the unit of circulation).
        Grey points in (C) and (D) are data averaged over late hold times (not shown in (A) and (B) for clarity).
        Data in (A), (B) and (D) are the mean $\pm$ standard error of the mean (s.e.m.) calculated from 10 to 25 measurements at each hold time.
    }
    \label{fig:grid_figure}
\end{figure}

Here we inject vortices into a uniform, planar BEC by dragging an optical grid formed by an array of laser beams through the atomic cloud~\cite{neely_observation_2010}, and observe the evolution of the resulting states.
Using velocity-selective Bragg scattering we obtain the sign and location of each vortex in the BEC~\cite{seo_observation_2017,methods}, and use a vortex classification algorithm~\cite{reeves_inverse_2013,valani_einstein-bose_2016}  to identify clustered vortices, dipole pairs and free vortices (Fig.~\ref{fig:example_images}).
By changing the size of the optical grid (Fig.~\ref{fig:grid_figure}), we change the initial spacing between vortices of opposite sign, which are preferentially shed from opposite sides of each obstacle, and thereby vary the kinetic energy of the vortices injected.
For the finest obstacles used, the resulting vortex distribution is dominated by dipole pairs, while for larger obstacles, clustered vortices form the majority, with the ratio changing monotonically with obstacle size (Fig.~\ref{fig:grid_figure}A).

\begin{figure}[ht]
    \includegraphics[]{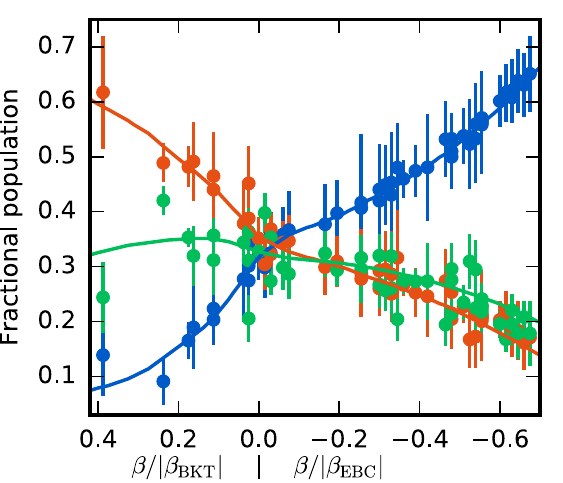}
    \centering
    \caption{\textbf{Temperature assignment.}
    The fractional populations of clustered (blue), dipole (red) and free (green) vortices averaged for each hold time after the sweep of each obstacle grid and placed on corresponding smoothed thermometer curves (solid lines) to assign an inverse temperature, $\beta$.
    The positive (negative) temperature axis is scaled by the critical temperature $|\beta_\mathrm{BKT}|$ ($|\beta_\mathrm{EBC}|$).
    The thermometry curves are generated by Monte Carlo simulations for 10 vortices, and the data placed at the least squares difference from the curves of both the cluster and dipole populations~\cite{methods}.
    Points show the mean populations $\pm$ s.e.m.
    } 
    \label{fig:average_temps}
\end{figure}

We analyze the vortex configurations using several independent measures.
The fractional populations $p_c$, $p_d$ and $p_f$ of clustered, dipole and free vortices, respectively, can be matched to thermometry curves generated by Monte Carlo simulations, to assign a temperature to a given state~\cite{methods,groszek_vortex_2018}.
Fitting ensemble averages (taken over repeated experiments with the same parameters) to the cluster and dipole thermometry curves shows good agreement with the expected behavior of all three populations (Fig.~\ref{fig:average_temps}).
As a secondary measure of the vortex clustering, we use the second order correlation function, $C_2$~\cite{methods}.
For a configuration dominated by dipoles (clusters), the sign of $C_2$ is negative (positive), and for a random configuration $C_2$ vanishes.
We numerically generate a spectrum of the incompressible kinetic energy per vortex for each configuration~\cite{methods,bradley_energy_2012}, which is integrated to give the total incompressible kinetic energy per vortex.
The cluster fraction, correlation function, temperature and energy per vortex all show an increasing trend as the grid is coarsened (Fig.~\ref{fig:grid_figure}A-D), which corresponds to an increase in long wavelength energy in the system (Fig.~\ref{extfig:gridspectra}).

\begin{figure*}[ht]
    \centering
    \includegraphics[]{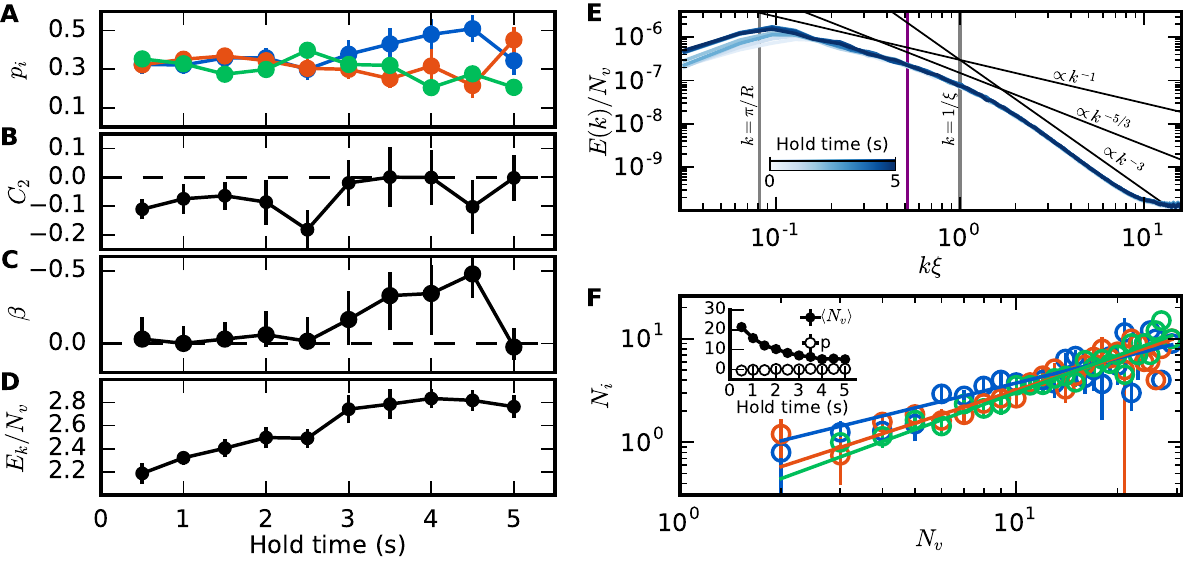}
    \caption{\textbf{Evidence of evaporative heating in the 6\,\boldmath$\mu$m grid data.}
    (\textbf{A}) Evolution of the classified vortex populations of clustered (blue), dipole (red) and free (green) vortices over 5\,s, starting from near equal weightings to more clustered states at later times.
    (\textbf{B}) Second order correlation function.
	(\textbf{C}) Inverse temperature of the vortices, $\beta$.
    (\textbf{D}) Total incompressible kinetic energy per vortex (in units of $\rho \kappa^2/4\pi$).
    (\textbf{E}) Incompressible kinetic energy spectrum (in units of $\rho \kappa^2/4\pi$).
    The increase in energy at wavenumbers on the order of the system size ($2R$) can be associated with the clustering of same-sign vortices.
    For wavenumbers $k\xi > 1$ the spectrum has a $k^{-3}$ scaling corresponding to the shape of a vortex core, while $k^{-1}$ corresponds to the far field of an isolated vortex~\cite{bradley_energy_2012}.
    A $k^{-5/3}$ scaling is expected for an energy cascade~\cite{kolmogorov_local_1941,kraichnan_inertial_1967}.
    The energy injection scale is $k = \pi/R_G$ (purple line), where $R_G =6.0\mu$m is the the semi-major axis of the grid.
    (\textbf{F}) Population of clusters ($N_c$), dipoles  ($N_d$) and free vortices  ($N_f$) as a function of total vortex number.
    The colored lines show power law fits to the data for each classification as described in the main text.
    The inset shows the mean of the vortex number ($\left<N_v\right>$) and polarization (p) as functions of hold time.
    Points in (A), (B), (D) and (E) are the mean $\pm$ s.e.m.
    }
    \label{fig:heating}
\end{figure*}

Time series data for each grid are shown in Figs.~\ref{fig:heating}, \ref{extfig:grid42}-\ref{extfig:grid115}.
For the 6\,$\mu$m grid (Fig.~\ref{fig:heating}), the distribution begins with approximately equal weightings of clusters, dipoles and free vortices, which corresponds to a temperature just to the positive side of infinity~\cite{groszek_vortex_2018}, and evolves to a state with a higher clustered fraction, and negative temperature (Fig.~\ref{fig:heating}A,C).
The correlation functions evolve from a negative average to a near-zero value (Fig.~\ref{fig:heating}B).
The incompressible kinetic energy spectrum shows a buildup of energy at scales on the order of the system size (Fig.~\ref{fig:heating}E), leading to an increase in the mean energy per vortex over time (Fig.~\ref{fig:heating}D).
These observations point towards the existence of an inverse energy cascade driving the system toward ordered Onsager vortex states. 
The scaling of the classified populations ($N_c$, $N_d$ \& $N_f$ for clusters, dipoles and free vortices, respectively) as a function of total vortex number $N_v$ (Fig.~\ref{fig:heating}F) are consistent with power laws, $N_c \propto N^{\alpha}_v$, $N_d \propto N^{\gamma}_v$ and $N_f \propto N^{\delta}_v$, that have been previously identified in numerical simulations.
The exponents obtained here are $\alpha = 0.80 \pm 0.09$, $\gamma = 1.06 \pm 0.08$ and $\delta = 1.19 \pm 0.07$, which compare well to those measured from the theory~\cite{groszek_vortex_2018}.

For an isolated system, energy can be supplied by the mechanism of `evaporative heating' of the vortices~\cite{simula_emergence_2014,groszek_vortex_2018}.
For an initially random distribution of vortices and antivortices, the annihilation of vortex-antivortex pairs leads to the removal of the `coldest' vortices---those contributing the least to the total incompressible kinetic energy associated with the vortex flow field.
The `hotter' remaining vortices rethermalize, with an increased mean energy per vortex.
This is directly analogous to the process of evaporative cooling, where the hottest particles are lost, resulting in a cooler distribution.
We propose that the aforementioned inverse cascade apparent in Fig.~\ref{fig:heating} is driven by this evaporative heating process.

In the configurations that begin with an appreciable clustered fraction, we do not observe significant evolution of the classified vortex fractions (Figs.~\ref{extfig:hot} and \ref{extfig:grid115}).
We believe that evaporative heating is suppressed in such configurations due to the small number of dipole and free vortices available to annihilate.
In fact, in these cases, the time evolution of the energy per vortex indicates that some cooling of the vortices occurs (see gray lines in Fig.~\ref{fig:grid_figure}).
As there is no significant evaporative heating in this regime, the cooling effects of phonons and the thermal background dominate the vortex thermodynamics.
We expect that for higher vortex numbers, with a similar cluster fraction, evaporative heating would in fact play a significant role, as the correspondingly higher number of dipole vortices remaining would allow for annihilations, as seen in numerical studies~\cite{simula_emergence_2014}.
This would drive the system further into the negative temperature regime, where potentially the condensation transition of the giant Onsager vortex clusters---known as the `Einstein--Bose' or `supercondensation' transition---could be induced~\cite{kraichnan_inertial_1967,valani_einstein-bose_2016,kraichnan_statistical_1975}.

The ability to measure statistical distributions arising from the dynamics of vortices will enable further studies of interacting quantum systems, such as experimental investigations of non-thermal fixed points~\cite{nowak_nonthermal_2012} in quantum turbulence and their relation to universal phenomena in far from equilibrium systems.
Indeed, the temporal evolution of the nearest neighbor inter-vortex spacings in our experiments (Fig.~\ref{extfig:non_thermal_fixed_point}) show signs of a predicted anomalous fixed point~\cite{karl_strongly_2017}.
Such non-equilibrium critical phenomena may have relevance in a wide range of systems, including inflation in the early universe~\cite{gasenzer_charge_2012}, quark-gluon plasma dynamics~\cite{arnold_qcd_2006}, and biologically motivated population models~\cite{reinhardt_nonequilibrium_2006}.

In summary, we have generated two-dimensional grid turbulence using a range of grid barriers of different dimensions.
The absolute temperature of the resulting vortex distributions was measured to range from positive through to negative values, with evidence of evaporative heating in their time evolution.
These results corroborate Onsager's statistical description of two-dimensional turbulence.
We note that the group of T.\ W.\ Neely has independently and simultaneously observed negative absolute temperature states of quantized vortices in a similar system, in the form of large persistent clusters~\cite{gauthier_giant_2018}.

We thank T.W.\ Neely, M.J.\ Davis, M.T.\ Reeves, G.\ Gauthier, A.S.\ Bradley, X.\ Yu, T.A.\ Bell, H.\ Rubinsztein-Dunlop and M.\ Baker for useful discussions.
We acknowledge financial support from the Australian Research Council via Discovery Projects DP130102321 (T.P.S., K.H.) and DP170104180 (T.P.S.). This research was also partially supported by the Australian Research Council Centre of Excellence in Future Low-Energy Electronics Technologies (project number CE170100039).

% \bibliography{references}

\section*{Methods}
\subsection*{Creation of a uniform, highly oblate BEC.}
We load $^{87}$Rb atoms, polarized in the $F=2$, $m_F = 2$ ground state, into an oblate trap formed by a 532\,nm Hermite--Gaussian HG$_{0,1}$ laser mode, and a quadrupole magnetic field gradient, as we have previously described~\cite{tempone-wiltshire_high_2017}.
We ramp on a linear bias field to 31.5\,G to push the magnetic quadrupole field zero above the atoms, reducing the radial trapping frequency to less than 1\,Hz, while maintaining a 15\,G\,cm$^{-1}$ gradient in the vertical direction to levitate the atoms against gravity.
Meanwhile, we project an optical in-plane trapping potential, also at 532\,nm, using a digital micromirror device (DMD), using the methods developed by Gauthier et al.~\cite{gauthier_direct_2016}.
In our apparatus, the DMD used is a Texas Instruments DLP LightCrafter evaluation module, with a 0.3 WVGA chipset.
Rather than using separate objective lenses for projection and imaging of the atoms, we use a single objective (Tech-Specialities, Inc.\ metallurgical near-infrared objective M Plan APO 20x with 0.35 numerical aperture, 20.2\,mm working distance and 3\,mm coverglass correction), with a dichroic mirror element to overlap the 532\,nm trapping light with the axis of the imaging system.
Our resolution is limited to approximately 2\,$\mu$m as we are imaging through a 6\,mm glass window.
We display a circular mask pattern on the DMD which forms a disc shaped potential in the trapping plane, with a radius of 39\,$\mu$m.
The magnification of the DMD is such that each pixel is approximately 0.6\,$\mu$m in the plane of the atoms, below the resolution of the imaging system.
We typically load partially condensed atoms into the HG mode, which are evaporatively cooled during the transfer to the final trap, with around $2\times10^5$ atoms remaining with a condensate fraction of at least 0.5.
We linearly reduce the intensity of both the HG mode and the DMD projection over 0.5\,s, until the axial trap frequency is 100\,Hz and the power from the DMD is just enough to maintain the radial confinement, as we experience higher atom loss rates at higher laser intensities.
The distribution of atoms in this trap appears uniform with a Gaussian edge of approximately 2\,$\mu$m.

\subsection*{Generation of grid turbulence}
We generate vortices by cycling through a sequence of patterns on the DMD containing grids of elliptical barriers, as depicted in Fig.~\ref{fig:grid_figure}, moving across the disc along the line defined by the directions of the Bragg spectroscopy laser beams.
Pairs of interleaved grids, moving in opposite directions across the BEC, are used to help suppress the generation of a net linear superfluid flow, thereby reducing the presence of center-of-mass motion of the cloud.
The longitudinal (along the direction of movement) width of each grid point element is fixed with a semi-minor axis of 3\,$\mu$m, while the semi-major axis of each point is varied from 4.2\,$\mu$m to 11.4\,$\mu$m, corresponding to the data points in Fig.~\ref{fig:grid_figure}.
The transverse spacing between ellipses in each grid corresponds to 3 semi-major axes, to ensure that when the counter propagating grids meet they do not touch.

We investigate the decay of the grid turbulence generated by the sweep by probing the vortex distributions at hold times from 0.5-5\,s in 0.5\,s increments.
Data presented in Fig.~\ref{fig:grid_figure} was averaged over data from 0.5-1.5\,s (early hold times) and 4-5\,s (late hold times).
We repeated the experiment 25 times for each hold time after the sweep of each obstacle grid.

\subsection*{Imaging vortices}
We use the Bragg spectroscopy technique introduced by Seo et al.~\cite{seo_observation_2017} to obtain information about the sign of each vortex.
Here we use light blue-detuned 6.6\,GHz from the $F=2$ to $F'=3$ transition in $^{87}$Rb.
We extinguish the HG and DMD traps, allowing the BEC to expand freely, leaving the magnetic field gradient on to prevent the atoms from falling from the imaging plane.
After 300\,$\mu$s we pulse the Bragg beams on for 600\,$\mu$s, with a detuning of $\delta_d/2\pi =900$\,Hz from the two-photon Bragg resonance.
The scattered parts of the cloud fully separate in 6\,ms, after which time the vortex cores in the condensate expand to a size resolvable by our imaging lens.
We use a 20\,$\mu$s pulse of resonant light to form an absorption image of the resulting condensate and Bragg scattered distributions of atoms.

\subsection*{Image processing}
We observe fringes in our imaging probe beam, which we cannot fully stabilize, believed to be due to multiple reflections between the windows of our vacuum chamber.
We minimize the severity of these fringes by focusing the imaging probe past the atoms, minimizing the overlap between the beam and its stray reflections.
To remove the effects of the motion of the remaining fringes, we use an eigenbasis method to generate flat field images~\cite{li_reduction_2007,ockeloen_detection_2010}.
A series of 200 flat field images, taken during similar runs of the experiment, are used to form an eigenbasis.
A masked copy of the absorption image is projected onto this basis to generate a frame that is a superposition of flat frame images best matching the light field away from the atoms.

\subsection*{Vortex distribution analysis}
We use the `blob detection' algorithm described by Rakonjac et al.~\cite{rakonjac_measuring_2016} to obtain an initial estimate of the locations of vortices from the optical density image.
Due to the turbulent nature of our vortex distributions, the algorithm is not as robust as the application in vortex lattices; the presence of sound waves and vortex dipole pairs too close together to distinguish can lead to both false positive and false negative detections, respectively.
To minimize these errors, we manually inspect the results of the algorithm and adjust if required.
We also reject images where the vortices are too hard to distinguish (this can occur, for example, if the fringes in the imaging probe are particularly strong, or if atoms become trapped outside the main nodal line of the HG mode, obscuring the BEC), or the atom number is significantly less than the average.

Once the locations of all vortices in the BEC have been identified, we use a point vortex model~\cite{simula_emergence_2014} to calculate the theoretical flow field due to each vortex and corresponding image vortex, for every combination of possible vortex signs.
We project these calculated flow fields onto the axis of the Bragg beams, on a grid matching the pixelation of our optical density images.
We create a differential signal from the images of the Bragg scattered components~\cite{seo_observation_2017}.
The sign of these projected flow fields is compared to that of the Bragg differential signal within 5\,$\mu$m of each vortex, and the vortex sign configuration with the highest number of matching pixels is chosen as the best configuration.
We find that, for each frame, only one of the possible configurations of vortices and antivortices provides quantitative agreement with the differential Bragg signals, and Fig.~\ref{fig:example_images} shows the corresponding qualitative agreement.
As an example, Fig.~\ref{extfig:sign_id} shows the number of matching pixels for each possible vortex configuration for the data shown in Fig.~\ref{fig:example_images}A, sorted from worst to best match.

With the sign and position of each vortex, we can calculate the second order correlation function, $C_2 = \frac{1}{2N_{v}} \sum_{i=1}^{N_v}\sum_{j=1}^{2}c_{ij}$, where $c_{ij} = 1\,(-1)$ if the circulation of the $j$th nearest neighbor of the $i$th vortex has the same (opposite) sign and $N_{v}$ is the total number of vortices detected.
We use the vortex classification algorithm outlined by Valani et al.~\cite{valani_einstein-bose_2016}.
This uniquely assigns each vortex as belonging to a cluster, dipole or as free, depending on the signs of its mutual nearest neighbors.
Correlation functions are calculated for each distribution, as described in the main text.

To calculate the incompressible kinetic energy spectrum we generate a density and velocity profile corresponding to the known vortex locations and signs.
For the density profile, we use a uniform disc with a $2\,\mu$m Gaussian blur, and the vortex core ansatz~\cite{bradley_energy_2012}
$$\chi(r) = \frac{r}{\sqrt{r^2 + (\xi\Lambda^{-1})^2}},$$
with a healing length $\xi = 1\,\mu$m, and $\Lambda = 0.825$.
We use a $280\,\mu$m $\times$ $280\,\mu$m grid with 1024 $\times$ 1024 pixels, to ensure that we capture wavenumbers $k = 2\pi/r$ ranging from beyond the trap scale to below the healing length.
The one-dimensional spectral density is then given by the radially averaging the Fourier transform of the density-weighted velocity field~\cite{bradley_energy_2012} into 512 $k$ bins and dividing by the number of vortices.
The total incompressible kinetic energy can then be calculated by integrating the spectrum over $k$.
A spectrum is generated for each hold time after the sweep of each obstacle grid size by averaging over the repeated experiments.

\subsection*{Thermometry}
The Boltzmann temperature of a vortex configuration is defined by its entropy, $S$, and energy, $E$, as $1/T \equiv \partial S / \partial E \equiv k_B\beta$, where $k_B$ is the Boltzmann constant.
To measure the vortex temperatures, we run Monte Carlo simulations~\cite{groszek_vortex_2018} to generate thermometer curves $p_c(\beta)$, $p_d(\beta)$ and $p_f(\beta)$ for the fractional populations of clustered vortices, dipole vortices and free vortices as a function of inverse temperature $\beta$.
We have scaled all positive (negative) inverse temperatures by the critical temperature $|\beta_\mathrm{BKT}|$ ($|\beta_\mathrm{EBC}|$) corresponding to the Berezinskii--Kosterlitz--Thouless (Einstein--Bose condensate) transition~\cite{kraichnan_two-dimensional_1980}.
For these simulations, we use a system of $N_v = 10$ point vortices, with zero net polarization, to approximate our experiments, which typically decay from around 20 vortices down to 5 in 5\,s.
Figure \ref{extfig:thermometry} demonstrates how these curves change as the vortex number is varied.
We smooth the temperature curves using a Savitzky--Golay filter, and take the global minimum in the root mean square difference, $R_{c,d}(\beta)=\sqrt{(p_c(\beta) - p_{c,\mathrm{meas}})^2 + (p_d(\beta) - p_{d,\mathrm{meas}})^2}$, between the thermometer curves and each ensemble average measurement of the cluster fraction, $p_{c,\mathrm{meas}}$, and dipole fraction, $p_{d,\mathrm{meas}}$, respectively, to correspond to the measured $\beta$ for each configuration.
We estimate the uncertainty in this measurement by finding $\beta_\mathrm{min}$ with $p_{c,\mathrm{meas}}-\mathrm{s.e.m.}$ and $p_{d,\mathrm{meas}}+\mathrm{s.e.m.}$, and $\beta_\mathrm{max}$ with $p_{c,\mathrm{meas}}+\mathrm{s.e.m.}$ and $p_{d,\mathrm{meas}}-\mathrm{s.e.m.}$

\subsection*{Data}
Time series data for each grid configuration are shown in Figs.~\ref{fig:heating}, \ref{extfig:grid42}-\ref{extfig:grid115}.
Each of these figures shows the evolution of the classified vortex populations, correlation function, inverse temperature and energy per vortex for the ensemble average over 10-25 repeated measurements at each hold time used (0.5\,s to 5.0\,s in 0.5\,s increments).
The averages of these values across the first three hold times (0.5\,s, 1.0\,s and 1.5\,s) are shown for each grid size in Fig.~\ref{fig:grid_figure} (`early hold times'), and the average energy and inverse temperature across the last three hold times (4.0\,s, 4.5\,s and 5.0\,s) are also shown (`late hold times').

\makeatletter 
\renewcommand{\thefigure}{S\@arabic\c@figure}
\makeatother
\setcounter{figure}{0}  

\begin{figure*}[h]
    \centering
    \includegraphics[]{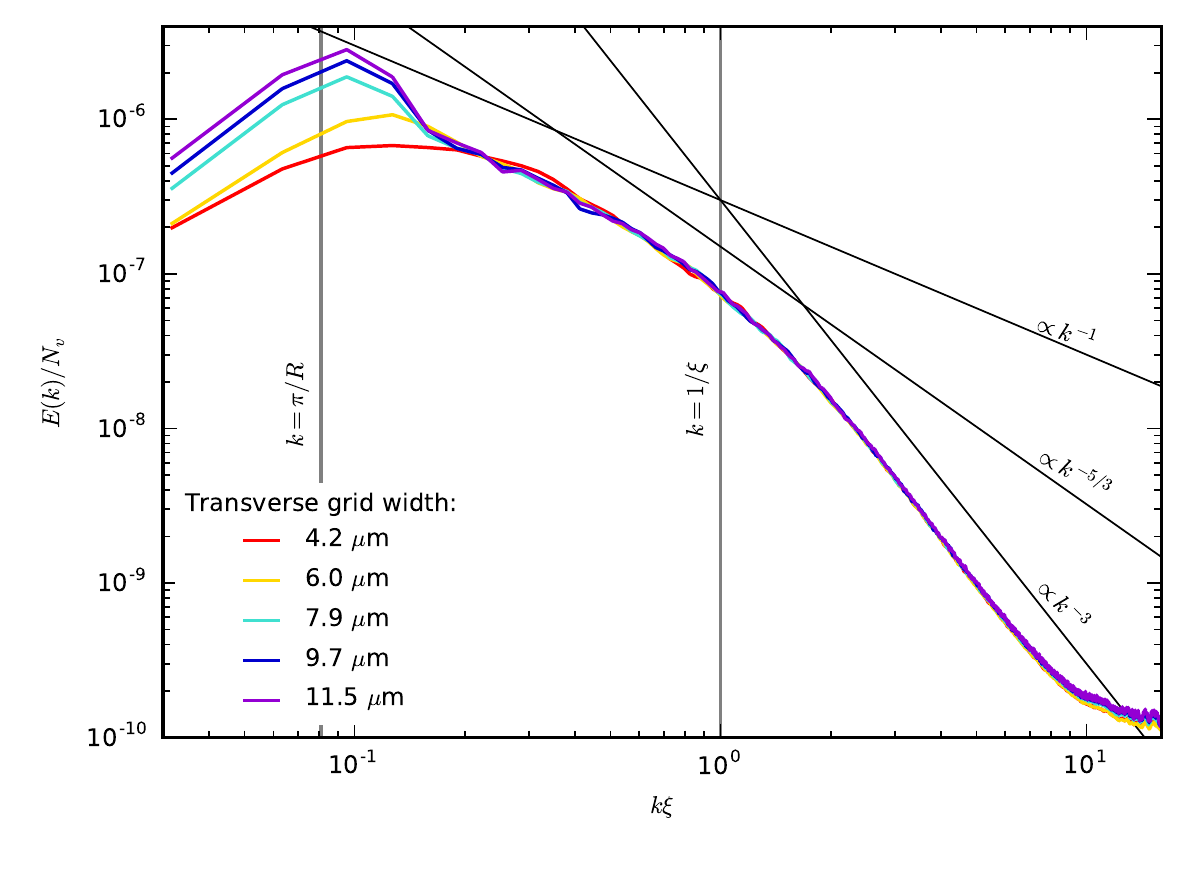}
    \caption{\textbf{Energy spectra for varying grid spacings.}
    Energy spectra (in units of $\rho \kappa^2/4\pi$) are averaged over early times for each grid size.
    As the grid becomes larger, an increase in energy is observed on scales on the order of the system size ($2R$).
    This is consistent with the observations of Fig.~\ref{fig:grid_figure}, indicating the clustering of same-sign vortices.
    }
    \label{extfig:gridspectra}
\end{figure*}

\begin{figure*}[h]
    \centering
    \includegraphics[]{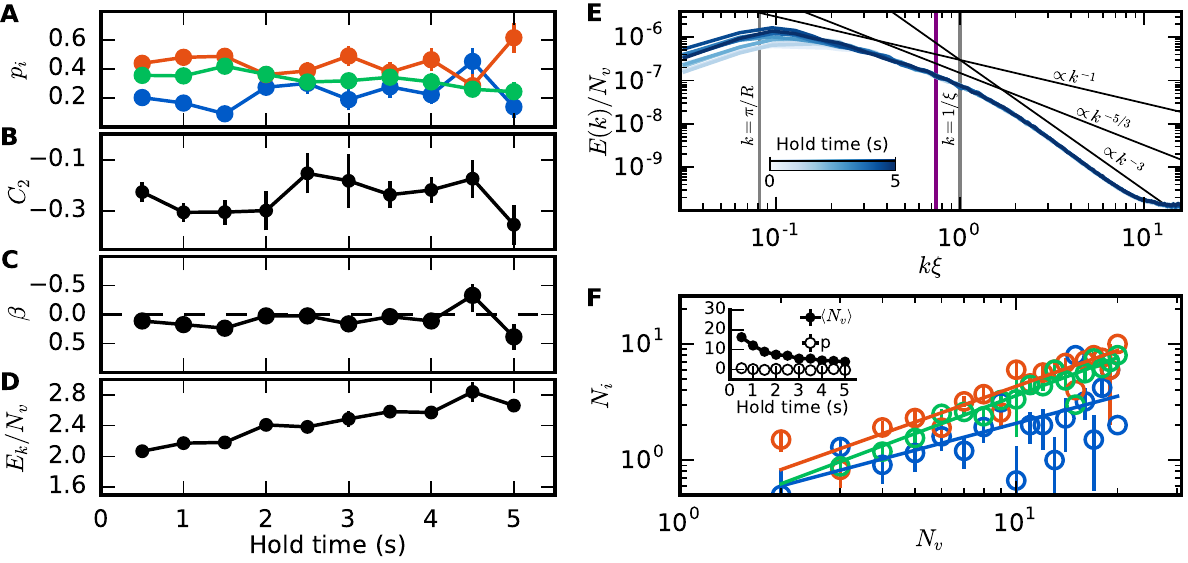}
    \caption{\textbf{4.2\,\boldmath$\mu$m grid dynamical evolution.}
    (\textbf{A}) Evolution of the classified vortex populations of clustered (blue), dipole (red) and free (green) vortices over 5\,s.
    (\textbf{B}) Second order correlation function.
	(\textbf{C}) Inverse temperature of the vortices, $\beta$.
    (\textbf{D}) Total incompressible kinetic energy per vortex (in units of $\rho \kappa^2/4\pi$).
    (\textbf{E}) Incompressible kinetic energy spectrum (in units of $\rho \kappa^2/4\pi$).
    The energy injection scale is $k = \pi/R_G$ (purple line), where $R_G =4.2\mu$m is the semi-major axis of the grid.
    (\textbf{F}) Population of clusters ($N_c$), dipoles  ($N_d$) and free vortices  ($N_f$) as a function of total vortex number.
    The colored lines show power law fits to the data for each classification: $N_c \propto N^{0.77 \pm 0.15}_v$, $N_d \propto N^{1.03 \pm 0.07}_v$ and $N_f \propto N^{1.09 \pm 0.06}_v$.
    The inset shows the mean of the vortex number ($\left<N_v\right>$) and polarization (p) as functions of hold time.
    Points in (A), (B), (D) and (E) are the mean $\pm$ s.e.m.
    }
    \label{extfig:grid42}
\end{figure*}

\begin{figure*}[h]
    \centering
    \includegraphics[]{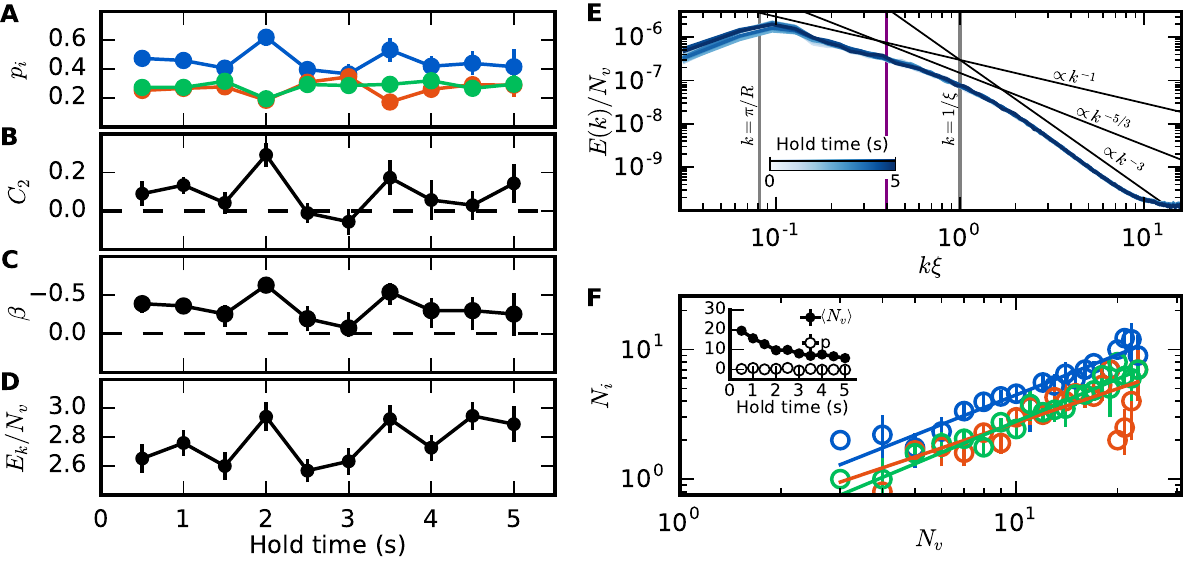}
    \caption{\textbf{7.9\,\boldmath$\mu$m grid dynamical evolution.}
    (\textbf{A}) Evolution of the classified vortex populations of clustered (blue), dipole (red) and free (green) vortices over 5\,s.
    (\textbf{B}) Second order correlation function.
	(\textbf{C}) Inverse temperature of the vortices, $\beta$.
    (\textbf{D}) Total incompressible kinetic energy per vortex (in units of $\rho \kappa^2/4\pi$).
    (\textbf{E}) Incompressible kinetic energy spectrum (in units of $\rho \kappa^2/4\pi$).
    The energy injection scale is $k = \pi/R_G$ (purple line), where $R_G = 7.9\mu$m is the semi-major axis of the grid.
    (\textbf{F}) Population of clusters ($N_c$), dipoles  ($N_d$) and free vortices  ($N_f$) as a function of total vortex number.
    The colored lines show power law fits to the data for each classification: $N_c \propto N^{1.03 \pm 0.10}_v$, $N_d \propto N^{0.87 \pm 0.12}_v$ and $N_f \propto N^{1.09 \pm 0.06}_v$.
    The inset shows the mean of the vortex number ($\left<N_v\right>$) and polarization (p) as functions of hold time.
    Points in (A), (B), (D) and (E) are the mean $\pm$ s.e.m.
    }
    \label{extfig:grid79}
\end{figure*}

\begin{figure*}[h]
    \centering
    \includegraphics[]{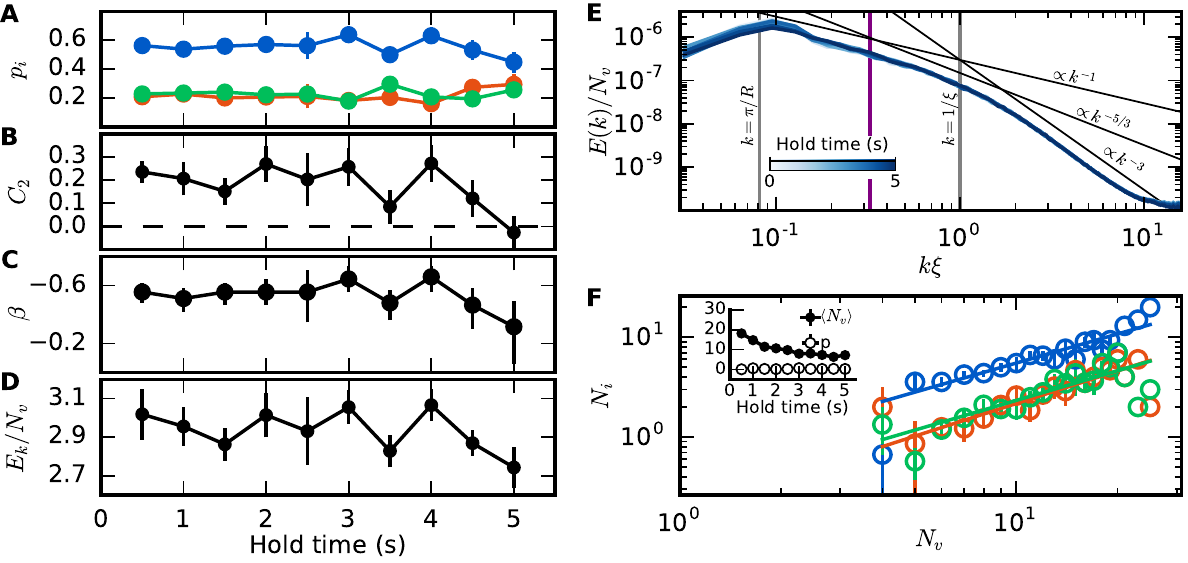}
    \caption{\textbf{9.7\,\boldmath$\mu$m grid dynamical evolution.}
    (\textbf{A}) Evolution of the classified vortex populations of clustered (blue), dipole (red) and free (green) vortices over 5\,s.
    (\textbf{B}) Second order correlation function.
	(\textbf{C}) Inverse temperature of the vortices, $\beta$.
    (\textbf{D}) Total incompressible kinetic energy per vortex (in units of $\rho \kappa^2/4\pi$).
    (\textbf{E}) Incompressible kinetic energy spectrum (in units of $\rho \kappa^2/4\pi$).
    The energy injection scale is $k = \pi/R_G$ (purple line), where $R_G =9.7\mu$m is the semi-major axis of the grid.
    (\textbf{F}) Population of clusters ($N_c$), dipoles  ($N_d$) and free vortices  ($N_f$) as a function of total vortex number.
    The colored lines show power law fits to the data for each classification: $N_c \propto N^{0.97 \pm 0.08}_v$, $N_d \propto N^{1.09 \pm 0.15}_v$ and $N_f \propto N^{0.99 \pm 0.12}_v$.
    The inset shows the mean of the vortex number ($\left<N_v\right>$) and polarization (p) as functions of hold time.
    Points in (A), (B), (D) and (E) are the mean $\pm$ s.e.m.
    }
    \label{extfig:hot}
\end{figure*}

\begin{figure*}[h]
    \centering
    \includegraphics[]{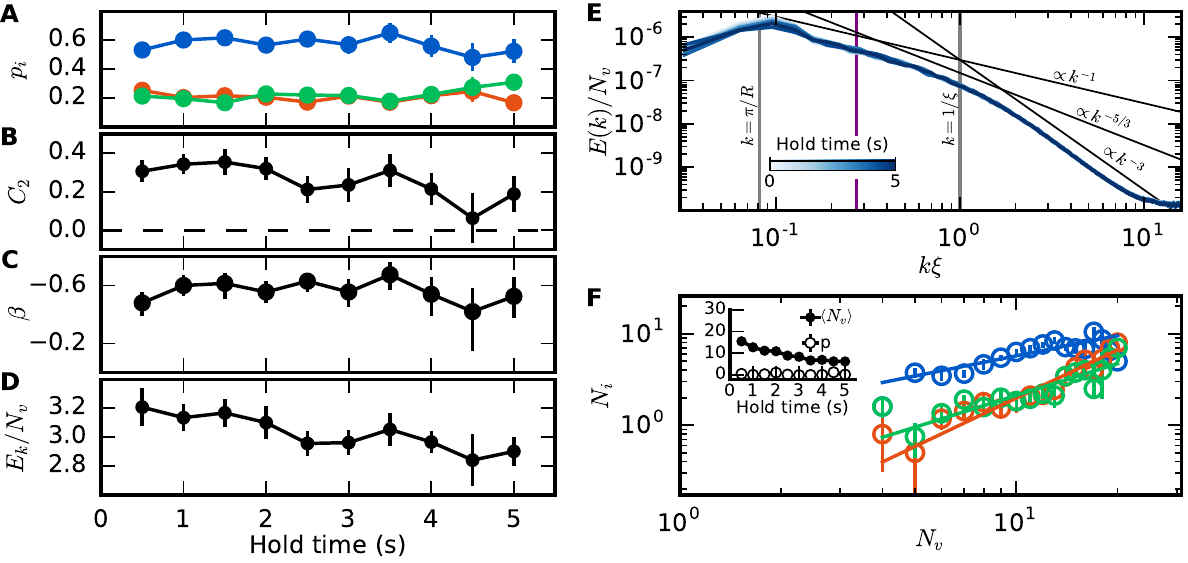}
    \caption{\textbf{11.5\,\boldmath$\mu$m grid dynamical evolution.}
    (\textbf{A}) Evolution of the classified vortex populations of clustered (blue), dipole (red) and free (green) vortices over 5\,s.
    (\textbf{B}) Second order correlation function.
	(\textbf{C}) Inverse temperature of the vortices, $\beta$.
    (\textbf{D}) Total incompressible kinetic energy per vortex (in units of $\rho \kappa^2/4\pi$).
    (\textbf{E}) Incompressible kinetic energy spectrum (in units of $\rho \kappa^2/4\pi$).
    The energy injection scale is $k = \pi/R_G$ (purple line), where $R_G =11.5\mu$m is the semi-major axis of the grid.
    (\textbf{F}) Population of clusters ($N_c$), dipoles  ($N_d$) and free vortices  ($N_f$) as a function of total vortex number.
    The colored lines show power law fits to the data for each classification: $N_c \propto N^{0.73 \pm 0.09}_v$, $N_d \propto N^{1.75 \pm 0.17}_v$ and $N_f \propto N^{1.15 \pm 0.16}_v$.
    The inset shows the mean of the vortex number ($\left<N_v\right>$) and polarization (p) as functions of hold time.
    Points in (A), (B), (D) and (E) are the mean $\pm$ s.e.m.
    }
    \label{extfig:grid115}
\end{figure*}

\begin{figure*}[h]
    \centering
    \includegraphics[]{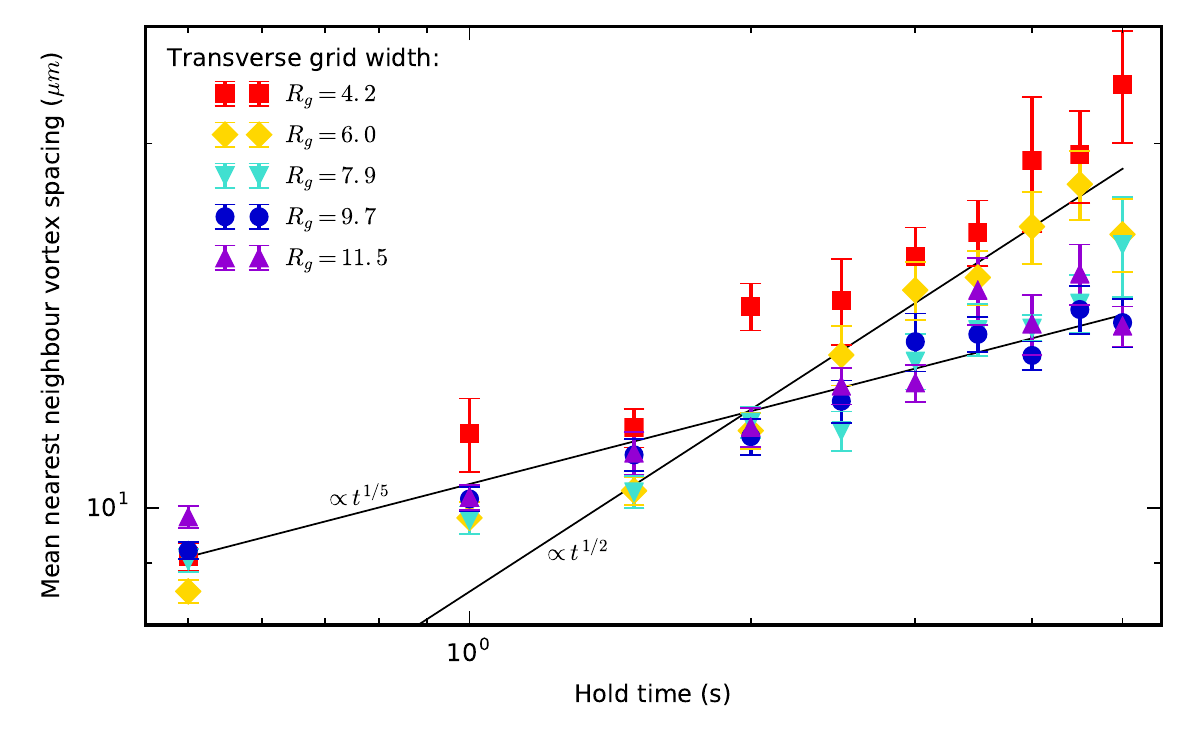}
    \caption{\textbf{Signs of an anomalous fixed point.}
    The mean nearest neighbor inter-vortex spacing plotted as a function of time for each grid.
    Error bars show s.e.m. over repeated experiments.
    In each case, at early times, the spacing appears to scale as $t^{1/5}$, which has been associated with an anomalous fixed point~\cite{karl_strongly_2017}.
    The data for the finer grids transitions to a $t^{1/2}$ slope, associated with the Gaussian fixed point, while this transition appears to be suppressed for the coarse grids.
    }
    \label{extfig:non_thermal_fixed_point}
\end{figure*}

\begin{figure*}[h]
    \centering
    \includegraphics[]{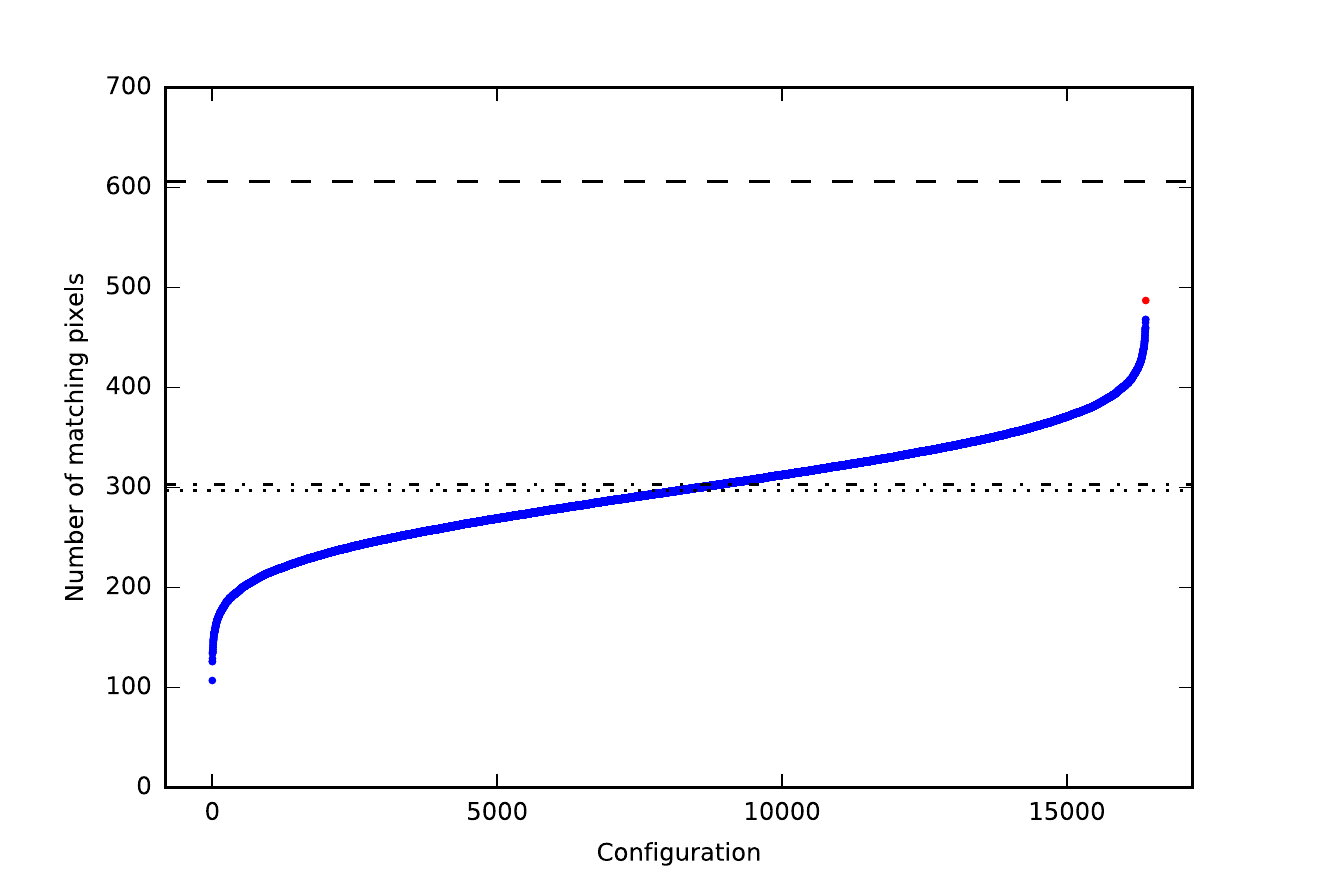}
    \caption{\textbf{Determining the best matching vortex sign configuration.}
    The number of pixels (within a 4 pixel radius of a vortex) where the sign of the Bragg signal matches the sign of the flow calculated by a point vortex model.
    In this case (data for top row of Fig.~\ref{fig:example_images}), there are 14 vortices, with $2^{14}$ possible sign configurations, and the signal has been compared on 606 pixels (dashed line).
    The average number of matching pixels across all possible configurations is 297 (dotted line), just below half of all pixels (dash-dot line), and the best match (red) was 487.
    }
    \label{extfig:sign_id}
\end{figure*}

\begin{figure*}[h]
    \centering
    \includegraphics[]{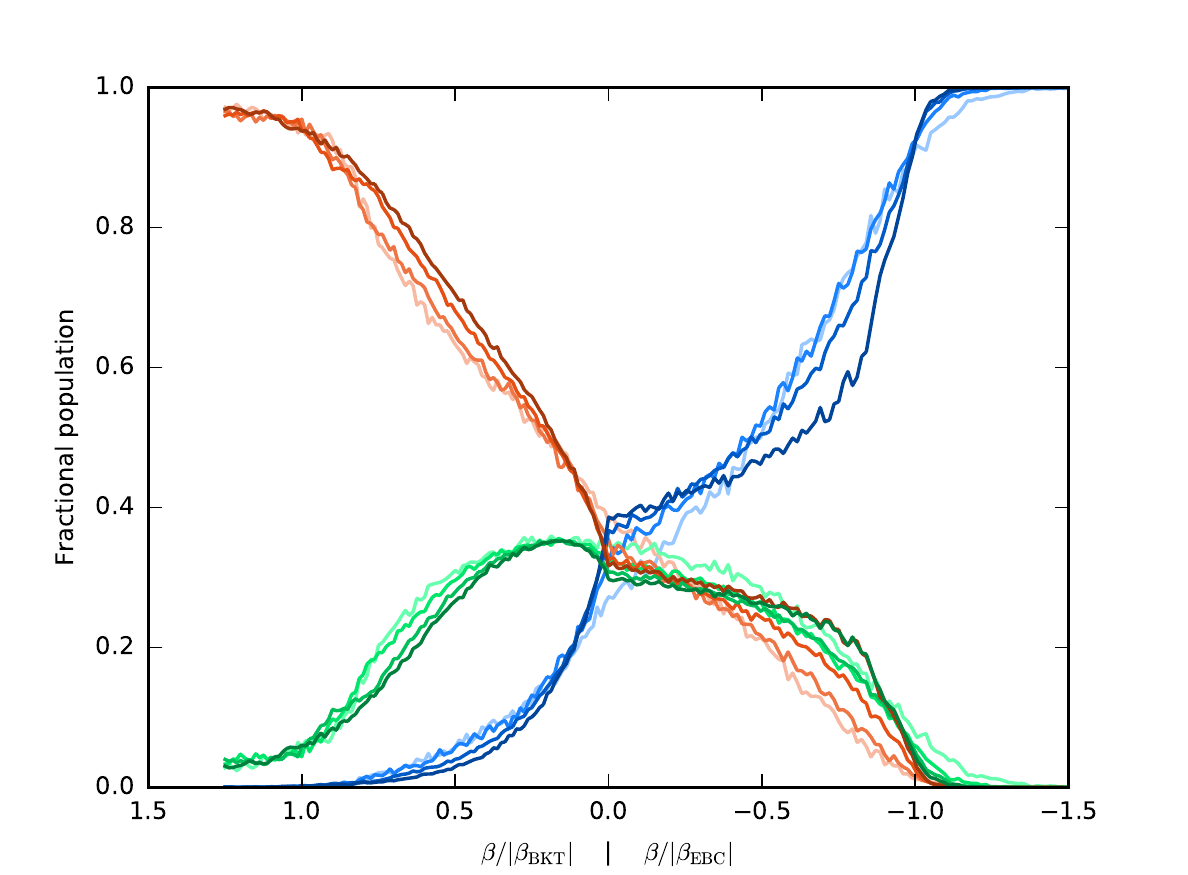}
    \caption{\textbf{Comparison of thermometry curves for varying vortex number.}
    Vortex thermometry curves $p_c(\beta)$ (blue), $p_d(\beta)$ (red) and $p_f(\beta)$ (green) are shown for $N_v$ = 50, 20, 10 and 6 (darkest to lightest).
    The curves are qualitatively similar, though the crossing of the populations is seen to shift slightly to more negative temperatures for lower vortex number.
    }
    \label{extfig:thermometry}
\end{figure*}

\end{document}